\documentclass[12pt,a4paper]{article}
\usepackage{ifpdf}
\usepackage{times}
\usepackage[scale={0.8,0.9},centering,includeheadfoot]{geometry}
\usepackage[shortlabels]{enumitem}
\usepackage{subcaption}
\usepackage{multicol}
\usepackage{graphics}
\usepackage{bm}
\usepackage{url}
\usepackage{rotating}
\usepackage{caption}
\usepackage{multicol}
\usepackage{multirow}
\usepackage{color}
\usepackage{graphicx}

\usepackage{listings}
\usepackage{pgf}

\usepackage{amsmath,amssymb,xspace,amsthm,bm}

\usepackage[colon,longnamesfirst]{natbib}

\usepackage{dcolumn}\newcolumntype{d}[1]{D{.}{.}{#1}}

\def\qedbox{\ifvmode\else\unskip\fi~\penalty10000
    \hfill{\large$\blacksquare$}}

\def\miscinfo#1#2{{\footnotesize\indent\textsc{#1: }\ignorespaces #2}}   

\definecolor{burgundy}{RGB}{144,0,32}
\definecolor{navyblue}{rgb}{0.0, 0.0, 0.5}
\definecolor{oceanboatblue}{rgb}{0.0, 0.47, 0.75}
\definecolor{mediumtealblue}{rgb}{0.0, 0.33, 0.71}

\begin{document}
\bibliographystyle{apalike}

\title{PC priors for residual correlation parameters in one-factor mixed models}

\author{
    Massimo Ventrucci, Daniela Cocchi \\
	Department of Statistical Sciences,\\    
    University of Bologna, 
    Via Delle Belle Arti 41, 40126, Bologna, Italy\\
    \\
    \and
    Gemma Burgazzi, Alex Laini\\
Department of Chemistry, Life Sciences and 
Environmental Sustainability,\\
    University of Parma,
    Parco Area delle Scienze, 33/A, 43124 Parma, Italy\\
    \\  
    }


\maketitle

\begin{abstract}
Lack of independence in the residuals from linear regression motivates the use of random effect models in many applied fields. 
We start from the one-way anova model and extend it to a general class of one-factor Bayesian mixed models, discussing several correlation structures for the within group residuals. All the considered group models are parametrized in terms of a single correlation (hyper-)parameter, controlling the shrinkage towards the case of independent residuals (iid). We derive a penalized complexity (PC) prior for the correlation parameter of a generic group model. This prior has desirable properties from a practical point of view: i) it ensures appropriate shrinkage to the iid case; ii) it depends on a scaling parameter whose choice only requires a prior guess on the proportion of total variance explained by the grouping factor; iii) it is defined on a distance scale common to all group models, thus the scaling parameter can be chosen in the same manner regardless the adopted group model. We show the benefit of using these PC priors in a case study in community ecology where different group models are compared.
\end{abstract}

\miscinfo{Keywords}{Bayesian mixed models, Group model; One-way anova; INLA; Intra-class correlation; Within group residuals.}\\
\miscinfo{Address for Correspondence}{Massimo Ventrucci, massimo.ventrucci@unibo.it}


\section{Introduction}

\subsection{Mixed models in community ecology}
The understanding of factors determining the distribution of organisms is a striking goal of 
community ecology \citep{heino2013environmental} and the key for forecasting the future trajectories of communities \citep{wisz2013role}. Modelling the organization and the evolution of natural communities is not an easy task, since the abundance and distribution 
of organisms depend on environmental drivers as well as on single taxon features and interactions among different taxa \citep{ovaskainen2017make,wisz2013role}. Several authors report high levels of unexplained variation, after considering the effect of environmental variables \citep*{lamouroux2004biological}. This residual variation is often ascribed to biotic (intra- and inter-specific) interactions, including both negative (like competition, predation and parasitism) and positive interactions (like mutualism and commensalism) which can play a crucial role in shaping communities.

A study in community ecology typically consists of observations of species abundance (biotic response variable) and environmental covariates (abiotic factors) collected at different locations and/or time points, through several sampling campaigns. At an early stage of the analysis ecologists' goal is to quantify the effect of the observed covariates, often assuming a linear regression model. Two alternative assumptions on the residuals are under examination:
\begin{itemize}
\item [(a)] residuals are iid; 
\item [(b)] residuals are correlated following some dependence structure. 
\end{itemize}
Assumption (a) means that the linear regression model including the observed covariates is the true model; in other words, what is not explained by covariates is iid noise. Assumption (b) means that residuals are not iid, thus there is unobserved heterogeneity in the data. It is crucial for ecologists to investigate the correlation pattern in the residuals, as this may reflect the presence of unobserved processes playing a role in shaping the underlying communities.

Mixed models are the most used tools for evaluating the two assumptions above, and have been extensively used in analyzing ecological data \citep{zuur-2009}.  
The popularity of mixed model in ecology is probably due to the fact that the effect of the observed covariates and unobserved processes can be neatly separated in the model. In its general formulation, a linear mixed model for a Gaussian response $\bm Y$ and covariates $\bm X$ is expressed as
\begin{equation*}
\bm Y = \bm X \bm \beta + \bm Z \bm b + \bm \epsilon, \quad ; \quad \bm b  \sim \mathcal{N}(\bm 0,\bm \Sigma_b) \quad ; \quad   \bm \epsilon \sim \mathcal{N}(\bm 0, \bm  I \sigma_e^2)
\label{eq:LMM}
\end{equation*}
where $\bm \beta$ are denoted as fixed effects and $\bm b$ as random effects, 
i.e.\ random variables with Gaussian distribution conditional on one or more variance (hyper-)parameter. The usual interpretation in ecology is that the $\bm \beta$'s account for variability in the data explained by \emph{observed} abiotic factors, while the $\bm b$'s account for sources of variability in $\bm Y$ driven by \emph{unobserved} abiotic or biotic factors \citep{warton-2015}. 


\subsection{One-way anova}

Let us consider one-way anova, which is the simplest mixed model case. Assume data are grouped according to the levels of a \emph{grouping factor}, with $y_{ij}$ being the response at unit $i=1,\ldots,m_j$ within group $j=1,\ldots,n$. 
The one-way anova model is:
\begin{eqnarray}
\label{eq:group}
y_{ij} & =& \alpha + \boldsymbol x_{ij}^{\textsf{T}} \bm \beta + b_j + \epsilon_{ij}   \quad \quad  i=1,\ldots,m_j  \quad j=1,\ldots,n,\\
b_j & \sim & \mathcal{N}(0, \sigma_{b}^2), \nonumber\\
\epsilon_{ij} & \sim &\mathcal{N}(0,\sigma^2_{\epsilon}), \nonumber
\end{eqnarray}
where $b_j$ and $\epsilon_{ij}$ are assumed as independent. The $b_j$'s are random effects quantifying \emph{group-specific} deviations from  
the global intercept $\alpha$. It is important to note that i) when $\sigma_b^2=0$ model (\ref{eq:group}) corresponds to 
$y_{ij} = \alpha + \boldsymbol x_{ij}^{\textsf{T}} \bm \beta + \epsilon_{ij}$, where 
only covariates matter and the rest is iid variation; 
ii) the random effects $b_j$'s and $\epsilon_{ij}$'s compete to capture the total residual variability, 
i.e. the variance of the terms $y_{ij} - \alpha - \boldsymbol x_{ij}^{\textsf{T}} \bm \beta$. 

By reparametrizing model (\ref{eq:group}) - see Section~\ref{sec:group} - it can be shown that the $b_j$'s induce correlation in residuals belonging to the same group; in this case, within group residuals are exchangeable. 
A possible ecological explanation for such type of unobserved heterogeneity is that members of the same group interact with each other or share some common features.

\subsection{Aim of the work}
Model (\ref{eq:group}) is an example of what in this paper is denoted as \emph{group model}, i.e. a model for the within group residual correlation structure. In particular, model (\ref{eq:group}) assumes an exchangeable group model. The first aim of this paper is to illustrate how group models beyond the exchangeable case can be constructed by extending formulation (\ref{eq:group}). The focus will be on models with one grouping factor, i.e. one-factor mixed models.  Comparing different group models can provide information on the main sources of heterogeneity in the data.  The need for reliable model comparison tools requires to build sensible priors for the correlation parameters in group models, which is the second purpose of this paper.

The usual practice in Bayesian analysis is to select independent prior distributions for 
the variance components $\sigma_b^2$ and $\sigma_{\epsilon}^2$ of model (\ref{eq:group}). The literature on priors for variance parameters is vast, see for instance \cite{gelman-2006}, and it is outside the scope of this work to give a comprehensive review of it. 
We note that common choices like conjugate priors lead to overfitting, as shown by several papers \citep{FruhwirthSchnatter-2010, FruhwirthSchnatter-2011, pcprior}.  

The priors we propose exploit two pieces of \emph{prior knowledge}, one regarding the structure of the model and the other concerning its variance components. The first piece of information is about the \emph{base model}, i.e. - applying the definition in \cite{pcprior} - the simplest model in the class of group models. Coherently with the assumptions (a) and (b) aforementioned, an ecologist would find that the natural base model for (\ref{eq:group}) is the one where $\sigma_b^2=0$, corresponding to assumption (a). We stress the fact that this information is certain and for free, as it simply reflects the fact that a mixed model is an extension of linear regression. The second piece of information - not for free and uncertain - is about the relative weight of the two variance components of the mixed model. The idea is that while an ecologist may have no opinion on the range of plausible values for $\sigma_{b}^2$ and $\sigma_{\epsilon}^2$, he/she may have an intuition on the relative importance of correlated residuals (controlled by $\sigma_b^2$) versus iid residuals (controlled by $\sigma_{\epsilon}^2$). This is essentially asking for a prior guess on the proportion of total (residual) variance explained by the grouping factor. 

In order to make good use of the knowledge about the base model we advocate priors that avoid overfitting by construction, i.e. priors that always give a chance to the iid base model to arise in the posterior, unless the data do require a more flexible one. Parsimony is a reasonable leading principle in
modelling ecological data, where unobserved heterogeneity plays a major role. 
It is therefore desirable to have flexible models that are able to shrink to simple ones. For building such priors we apply the \emph{Penalized complexity prior} framework by \cite{pcprior}.
 
In order to exploit the second piece of information, we abandon the idea of selecting independent priors for the variance components. Instead, we exploit a common reparameterization of model (\ref{eq:group}) and define a prior on the intra-class correlation (ICC) paramater, i.e. the proportion of total residual variance explained by the grouping factor. The resulting PC prior depends on a scaling parameter that can intuitively be elicited based on a prior opinion on the ICC. Importantly, this scaling can be used in general for any group model because PC priors are invariant over reparameterization. 

The plan of the paper is as follows. In Section~\ref{sec:motivating} the motivating example is described. In Section~\ref{sec:group} group models are presented, distinguishing between the cases of exchangeable and structured residuals. The PC prior for the correlation parameter in a generic group model is derived in Section~\ref{sec:PCpriors}, with additional results regarding the balanced design case. An illustration of the proposed group models and PC priors is given in Section~\ref{sec:application}, where we also describe a strategy to perform group model comparison by means of the Bayes factor. The paper closes with a discussion in Section~\ref{sec:discussion}.


\section{Motivating example}
\label{sec:motivating}

The motivating example for this work concerns a study on macroinvertebrate communities from data 
collected in six sampling campaigns carried out in three different streams, tributaries of the Po River (Northern Italy): 
Nure Stream, Parma Stream and Enza Stream. For each river a sampling area was selected and sampled twice, once in summer and once in winter. The spatial design included $50$ random points in each area, aligned along several transects. 
At each point, abundance of macroinvertebrates (response) and environmental covariates such as flow velocity, water depth, substrate composition and benthic organic matter were recorded. 
 
The application goal is to investigate the role of the environmental covariates and the presence of small scale processes within macroinvertebrate communities. In Section~\ref{sec:application} we propose different group models for the residuals, as an exploratory analysis to understand the main sources of unobserved heterogeneity. If we consider \texttt{campaign} as grouping factor, data are grouped in $n=6$ group/campaigns having $m=50$ observations. If we consider \texttt{transect} as grouping factor, data are grouped in $n=38$ groups/transects, each having a varying number (between $7$ and $10$) of irregularly-spaced observations. 


\section{Group models}
\label{sec:group}
We distinguish between two broad classes of models for the within group residuals: exchangeable residuals and structured residuals.
For the latter, we discuss in detail two group models, namely the autoregressive of order 1 (AR1) process and its continuous version known as the Ornstein Uhlenbeck (OU) process; these are particularly useful for the case study considered in this paper, but are also relevant in general applications within ecology where observations are often taken at different time points. Other group models can be 
constructed following the same idea illustrated here.  

\subsection{Exchangeable residuals}
\label{sec:group_exch}
Let us reparametrize (\ref{eq:group}) as 
\[y_{ij} = \alpha +  \bm x_{ij}^{\textsf{T}} \bm \beta+ \theta_{ij} \quad \quad  i=1,\ldots,m_j  \quad j=1,\ldots,n, \]
where $\theta_{ij}=b_j+\epsilon_{ij}$ is the residual at unit $i$ within group $j$. It follows that
\[Corr(\theta_{ij}, \theta_{hj})=\frac{\sigma^2_b}{\sigma^2_b+ \sigma_{\epsilon}^2} \quad \quad  i,h \in \{1,\ldots,m_j\} \quad \forall j.\] 
This means that the distribution of the residuals within group $j$, $\bm\theta_j = (\theta_{1j}, \ldots, \theta_{mj})^T$, is unchanged under permutation of the indexes $1,\ldots,m$, leading to the \emph{exchangeable} model
\begin{eqnarray}
\bm \theta_j & \sim  & \mathcal{N}(0, \sigma^2 \bm R_j(\rho)) \quad \forall j,
\label{eq:group-reparam}
\end{eqnarray}
where $\sigma^2=\sigma^2_b+ \sigma_{\epsilon}^2$ is the total variance and the correlation matrix is
\begin{equation}
\boldsymbol{R}_j(\rho) = 
\begin{bmatrix}
  1 & {\rho} & \cdots  & \cdots      & {\rho}   \\
{\rho} & 1  & \ddots &  &  \vdots \\
   \vdots & \ddots  & \ddots&   \ddots&  \vdots   \\
   \vdots &    & \ddots & 1   &  \rho \\
   {\rho} &  \cdots    & \cdots &
     {\rho} & 1 \\
\end{bmatrix}.
\label{eq:compound-matrix}
\end{equation}
Following the constrained interpretation \citep{verb-2003}, where the variance components are restricted to be non negative, 
the correlation parameter $\rho=\sigma^2_b/(\sigma^2_b+ \sigma_{\epsilon}^2)$ is the proportion of total (residual) variance explained by the grouping factor, i.e. the ICC. We use notation $\boldsymbol{R}_j(\rho)$ to emphasize that the correlation matrix depends on $\rho$. Note that if design is balanced then $m_j=m, \forall j$, hence $\bm{R}_j(\rho)=\bm{R}(\rho), \forall j$.

Matrix (\ref{eq:compound-matrix}) implies that residuals within each group are mutually correlated, with correlation parameter equal to $\rho$. We note that for any $\sigma^2>0$, $\rho=0$ identifies the linear regression model, thus $\rho$ is the parameter responsible for the shrinkage towards the base model. In Section~\ref{sec:PCpriors} we introduce a PC prior for $\rho$ depending on a scaling parameter that can be specified in a very intuitive manner, using a prior statement about the ICC. 

\subsection{Structured residuals}

\subsubsection{Autoregressive of order one}
\label{sec:group_ar1}
The AR1 process is often used to model correlation over time, when observations are taken at regularly-spaced time points (e.g. days, weeks, etc). Assume the model
\begin{eqnarray}
y_{ij} & = & \alpha + \boldsymbol x_{ij}^{\textsf{T}} \bm \beta + b_{ij} + \epsilon_{ij}  \quad \quad  i=1,\ldots,m_j  \quad j=1,\ldots,n, \nonumber \\
b_{1j} & \sim & \mathcal{N}(0, \sigma_{b}^2)  \quad \quad ; \quad \quad  b_{ij}   \sim  \mathcal{N}(\tilde\rho b_{{i-1}, j}, \sigma_I^2)  \quad i = 2,\ldots,m_j, \nonumber \\
\epsilon_{ij} & \sim &\mathcal{N}(0,\sigma^2_{\epsilon}), \nonumber
\end{eqnarray}
which defines an AR1 process on $b_{ij}, i=1,\ldots,m$, with $\sigma_I^2$, $\tilde\rho$ and $\sigma_{b}^2 = \frac{\sigma_I^2}{1-\tilde\rho^2}$ being, respectively, the innovation variance, the lag-one correlation and the marginal variance of the process.

Analogously to the exchangeable case, by reparameterizing $\theta_{ij} = b_{ij} + \epsilon_{ij}$, it follows that:
\[Corr(\theta_{ij}, \theta_{hj}) = \frac{\sigma^2_{b}}{\sigma^2_{b} + \sigma^2_{\epsilon}} \tilde\rho^{|i-h|} \quad  \quad i,h \in \{1,\ldots,m_j\} \quad \forall j. \]
In compact notation the group model can be rewritten as in Eq. (\ref{eq:group-reparam}) with total variance $\sigma^2 = \sigma^2_{b} + \sigma^2_{\epsilon}$ and correlation matrix,
\begin{equation}
\boldsymbol{R}_j(\rho) = 
\begin{bmatrix}
  1 & {\rho} & \rho^{2} & \cdots  &    \cdots    & \rho^{m-1} & {\rho}^{m}   \\
{\rho} & 1  & {\rho} & \rho^2 &\ldots    & \rho^{m-2}& {\rho}^{m-1}   \\
\rho^2 &   \rho  & 1   &   \rho & \rho^2  &  \cdots  & \rho^{m-2}  \\
   \vdots &   \ddots &  \ddots & \ddots & \ddots  & \ddots  & \vdots  \\
  \rho^{m-2}  & \cdots &  \rho^2 & \rho & 1 & \rho &  \rho^2 \\
  {\rho}^{m-1} &  \rho^{m-2} & \ldots & \rho^2 & \rho & 1 & \rho\\
      {\rho}^m &   {\rho}^{m-1}    & \cdots & \cdots & \rho^2& {\rho} & 1
\end{bmatrix},
\label{eq:ar1-matrix}
\end{equation} 
where the correlation parameter at generic lag $|i-h|$ is equal to $ \rho^{|i-h|}=\frac{\sigma^2_{b} }{\sigma^2_{b} + \sigma^2_{\epsilon}}\tilde\rho^{|i-h|}$.

This group model implies that within group residuals are correlated, with correlation structure driven by the ordering of the observations. 
We note that, analogously to the exchangeable case, $\rho=0$ identifies the linear regression model, for any $\sigma^2>0$. In Section~\ref{sec:PCpriors} we will describe a PC prior for $\rho$ whose scaling parameter can be chosen in the same intuitive way as in the exchangeable case.

\subsubsection{Ornstein Uhlenbeck}
\label{sec:group_ou}
The OU process is the continuous version of the AR1 and is appropriate when the observations are not equally-spaced.  The model is conceptually the same as the AR1 but reparameterized to account for the distances between locations. Let us assume $\delta_{ih}$ is the distance between observations $i$ and $h$, the correlation is 
\[Corr(\theta_{ij},\theta_{hj}) = \frac{\sigma^2_{b}}{\sigma^2_{b} + \sigma^2_{\epsilon}}\exp(-\delta_{ih}\phi), \quad \quad \phi >0. \]
This model can also be rewritten as in Eq. (\ref{eq:group-reparam}) with total variance $\sigma^2 = \sigma^2_{b} + \sigma^2_{\epsilon}$ and correlation matrix as in (\ref{eq:ar1-matrix}), with correlation between $i$ and $h$ equal to $ \rho_{i,h}=\frac{\sigma^2_{b}}{\sigma^2_{b} + \sigma^2_{\epsilon}} \exp(-\delta_{ih}\phi)$, thus the only difference w.r.t. the AR1 case is that the correlation parameter is expressed in log scale, $\phi = -\log(\rho)$. The OU process has good computational properties due to its sparse tridiagonal precision matrix, whose structure is detailed in \cite{finley-2009}. We use the OU process to model residual correlation when observations are randomly located along transects, as it is the case in our motivating example. 

Note that, for any $\sigma^2>0$, the linear regression model is achieved in the limit for $\phi \to \infty$. In Section~\ref{sec:PCpriors} we describe a PC prior for $\phi$ whose parameters can be specified in the same intuitive way as in the exchangeable case.


\section{PC priors for group models}
\label{sec:PCpriors}

In each of the group models presented above we have two (hyper-)parameters that need to be assigned a prior, the marginal variance $\sigma^2$ and the correlation $\rho$. 
What is relevant for us is to derive the PC prior for $\rho$, as this is the only parameter responsible for the shrinkage to the iid base model. 


A PC prior is defined in \cite{pcprior} as an exponential distribution on a \emph{distance} scale, measuring model complexity w.r.t. the base model. \cite{pcprior} argue that,  because the mode of the exponential is at distance $0$ (i.e. at the base model),  PC priors guard against overfitting by construction as they always give non-zero probability to a neighbourhood of the base model. This argument is based on an informal definition of an \emph{overfitting prior} as a prior density that is zero at the base model \citep{pcprior}; the idea is that an overfitting prior may drag the posterior away from the base model even when the latter is the true model; on the contrary a prior that contracts to the base model prevents overfitting by default. Several papers have confirmed tendency of PC priors to prevent overfitting via simulation studies \citep{Fuglstad:2017, klein-2015, ventrucci-rue-2016}.

Another property of PC priors which is exploited in the context of this paper is that they are invariant over reparameterization, as they are defined on a distance scale (instead of the original scale of the parameter), then translated in the scale of the original parameter by the change of variable rule.

\subsection{The PC prior for $\rho$}
With no loss of generality we present the steps to construct the PC prior for $\rho$ assuming $\sigma^2=1$; for a full discussion of the principles underpinnning the PC prior framework see \cite{pcprior}. 

Let $\bm \theta=\left( \bm \theta_1, \ldots, \bm \theta_n\right)^{\textsf{T}}$ denote the vector of residuals from all groups, $\pi(\bm \theta)$ the flexible model and $\pi_0(\bm \theta)$ the base model. 
The base model corresponds to $\rho=0$, as
\[
\pi_0(\bm \theta) = \mathcal{N}(\bm 0, \bm C_0) \quad ;\quad \bm C_0=\bm I_M,
\] 
where $\bm C_0$ is the identity matrix of dimension $M=\sum_{j=1}^{n} m_j$. The flexible model is when $\rho>0$, hence
\[
\pi(\bm \theta) = \mathcal{N}(\bm 0, \bm C) \quad ;\quad \bm C=\text{diag}\left\{\bm R_1(\rho), \ldots, \bm R_n(\rho)\right\},
\] 
where $\bm C$ is a block diagonal matrix containing all the within group correlation matrices.

The first step is computation of the Kullback Leibler divergence (KLD, \cite{kld-1951}) between the flexible and the base model, 
\begin{equation*}
\text{KLD}(\pi||\pi_0)  = \int \pi(\boldsymbol\theta) \log\left(\frac{\pi(\boldsymbol\theta)}{\pi_0(\boldsymbol\theta)}\right) d\boldsymbol\theta.\end{equation*}
In our case $\pi$ and $\pi_0$ are zero-mean multivariate normal densities of dimension $M$ with covariances $\bm C$ and $\bm C_0$,  respectively, thus the KLD simplifies to:
\begin{equation*}
\text{KLD}(\pi||\pi_0) =  \frac{1}{2}\left[\text{trace}\left( \bm C_0^{-1}\bm C\right)-M-\log\left(\frac{| \bm C|}{|\bm C_0|}\right)\right],
\end{equation*}
where notation $|\cdot|$ indicates the matrix determinant.
Given that $\bm C$ has block diagonal structure, we obtain:
\begin{equation*}
\text{KLD}(\pi||\pi_0)  = -\frac{\sum_{j=1}^n \log\left(|\bm{R}_j(\rho)|\right)}{2}.
\end{equation*}
For mathematical convenience, the distance from the base model is expressed as $\sqrt{2{\text{KLD}}(\pi_1||\pi_0)}$, which gives 
\begin{equation}
d(\rho) =  \sqrt{-\sum_{j=1}^n \log\left(|\bm{R}_j(\rho)|\right)}.
\label{eq:distance}
\end{equation}
From Eq. (\ref{eq:distance}) we see that the distance from the base model is a function of $\rho$ which takes values in the interval $[0,\infty)$; it is $0$ at the  base model ($\rho=0$) and goes to $\infty$ as $\rho \to 1$. 

The next step requires to specify an exponential distribution on $d(\rho)$ with rate $\lambda$,
\begin{equation}
\pi(d(\rho)) = \lambda \exp\left(-\lambda d(\rho) \right), \quad   \quad \lambda>0.
\label{eq:pcprior}
\end{equation}
Here $\lambda$ plays the role of a scaling parameter, controlling the degree of penalty for deviating from the base model. The larger $\lambda$, the stronger the penalty for deviating from the base model, at prior. Finally, the PC prior for $\rho$ is derived by the change of variable rule:
\begin{eqnarray}
	\pi(\rho) & = & \lambda \exp\left(-\lambda d(\rho)\right) \left|\frac{\partial d(\rho)}{\partial \rho}\right| \nonumber\\
& = & \sum_{j=1}^n\left( {|\bm R_j(\rho)|}^{-1} \frac{\partial |\bm R_j(\rho)|}{\partial \rho}\right)  \frac{\lambda}{2 d(\rho)}  \exp\left(-\lambda d(\rho)\right)  \quad \quad 0 \leq\rho < 1. 
\label{eq:pcprior_rho}
\end{eqnarray}

The PC prior in Eq. (\ref{eq:pcprior_rho}) depends on $\rho$ through the determinant of the within group $j$ correlation matrix $|\bm R_j(\rho)|$ and its derivative $\frac{\partial |\bm R_j(\rho)|}{\partial \rho}$. Therefore, one can derive analytically (or compute numerically) the PC prior for different group models by just plugging-in the determinant and its derivative in (\ref{eq:pcprior_rho}). In order to implement the PC prior in our model we need to choose $\lambda$ in Eq. (\ref{eq:pcprior_rho}); we postpone the discussion on how to choose it to Section~\ref{sec:lambda}. 

\subsection{The balanced design case}
\label{sec:PCpriors-balanced}
If design is balanced, PC priors for all the group models of Section~\ref{sec:group} can be derived analytically. In the unbalanced case, closed form expressions are more involved and will not be presented here; a practical solution for unbalanced designs is to evaluate the PC prior numerically. Below we report the PC priors for the correlation parameter in each group model of Section~\ref{sec:PCpriors}, for the balanced case. See Appendix~\ref{sec:app_proofs} for the mathematical details.

\subsubsection{Exchangeable residuals}

In the exchangeable case, $\rho$ is the within group correlation. 
The distance is
\begin{equation*}
d(\rho)=\sqrt{-n\log\left( (1+(m-1)\rho) (1-\rho)^{m-1}\right)} \quad \quad 0 \leq\rho < 1.
\label{eq:d}
\end{equation*}
The PC prior is
\begin{equation}
	\pi(\rho)=\frac{m-1}{2}\left(\frac{1}{1-\rho} - \frac{1}{1+(m-1)\rho}\right) \frac{\lambda'}{\sqrt{-\log\left(|\bm{R}(\rho)|\right)}} \exp(-\lambda' \sqrt{-\log\left(|\bm{R}(\rho)|\right)}),
\label{eq:pcprior_rho_exch}
\end{equation}
where $|\bm R(\rho)| = (1+(m-1)\rho)(1-\rho)^{m-1}$ and $\lambda' = \lambda \sqrt{n}$. 

\subsubsection{Structured residuals: AR1}

In the AR1 case, $\rho$ is the lag-one correlation parameter. The distance is
\[
d(\rho) =  \sqrt{n(1-m)  \log(1-\rho^2)}  \quad \quad 0 \leq \rho < 1.
\]
The PC prior is
\begin{equation}
\pi(\rho)  = \frac{\rho(m-1)}{1-\rho^2}\frac{\lambda'}{\sqrt{-\log\left(|\bm{R}(\rho)|\right)}} \exp(-\lambda' \sqrt{-\log\left(|\bm{R}(\rho)|\right)}),
\label{eq:pcprior_rho_ar1}
\end{equation}
where $|\bm R(\rho)| = (1-\rho^2)^{m-1}$ and $\lambda' = \lambda \sqrt{n}$. 

\subsubsection{Structured residuals: OU}

In the OU case, $\phi = -\log(\rho)$ is the lag-one correlation parameter expressed in the log scale. The distance is
\[
d(\phi) =  \sqrt{n(1-m)  \log(1-\exp(-2\phi))} \quad \quad  \phi >0.
\]
The PC prior is
\begin{equation}
\pi(\phi)  = \frac{(m-1)\exp(-2\phi)}{1-\exp(-2\phi)}\frac{\lambda'}{\sqrt{-\log\left(|\bm{R}(\phi)|\right)}} \exp(-\lambda' \sqrt{-\log\left(|\bm{R}(\phi)|\right)}),
\label{eq:pcprior_rho_ou}
\end{equation}
where $|\bm R(\phi)| = (1+\exp(-2\phi))^{m-1}$ and $\lambda' = \lambda \sqrt{n}$. 

\subsection{Choice of $\lambda$}
\label{sec:lambda}
The degree of informativeness of the PC prior in Eq. (\ref{eq:pcprior}) can be managed through $\lambda$, that defines the ``prior distance'' from the base model. \cite{pcprior} proposed to select $\lambda$ through the following rule: set values $U$ and $a$ such that $\mathbb{P}(\rho < U)=a$. By working out the cumulative distribution function of $\pi(\rho)$ we have,
\begin{equation}
\mathbb{P}(\rho < U)= \int_{0}^U \lambda \exp(-\lambda d(\rho))\left| \frac{\partial d(\rho)}{\partial \rho} \right| = 1-\exp(-\lambda d(U)) = a. 
\label{eq:lambda0}
\end{equation}
Solving Eq.~(\ref{eq:lambda0}) for $\lambda$ we obtain 
\begin{equation}
\lambda = -\log(1-a) / d(U),
\label{eq:lambda}
\end{equation}
hence, for the PC priors in Section~\ref{sec:PCpriors-balanced}, $\lambda'=-(\sqrt{n}\log(1-a)) / d(U)$.
Note that $\lambda$ in Eq. (\ref{eq:lambda}) can be computed in a generic group model (e.g. exchangeable, AR1, OU) by just plugging-in the associated distance function evaluated at $U$. 

\begin{figure}
\centerline{
\includegraphics[angle=270,width=.5\textwidth]{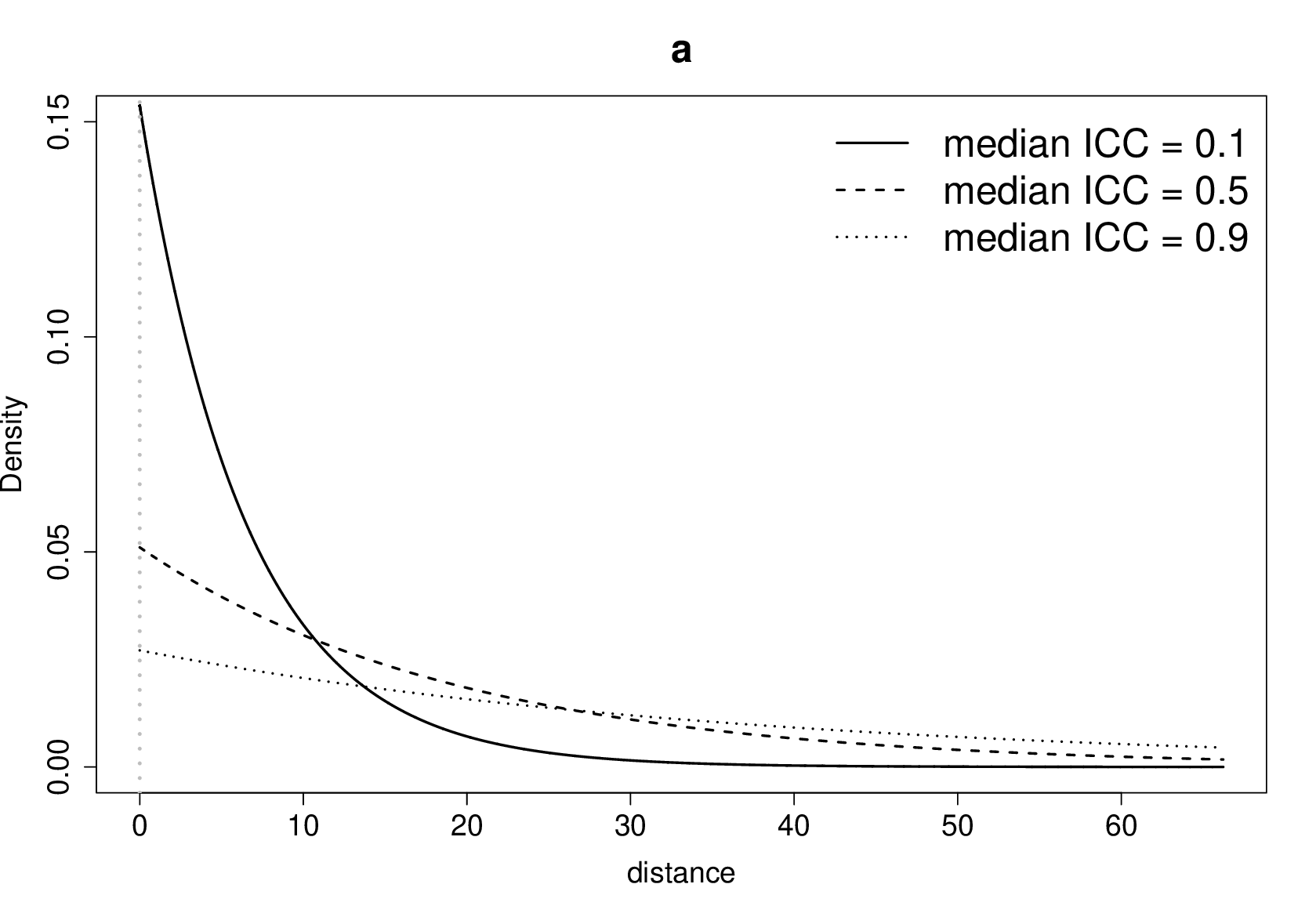}
\includegraphics[angle=270,width=.5\textwidth]{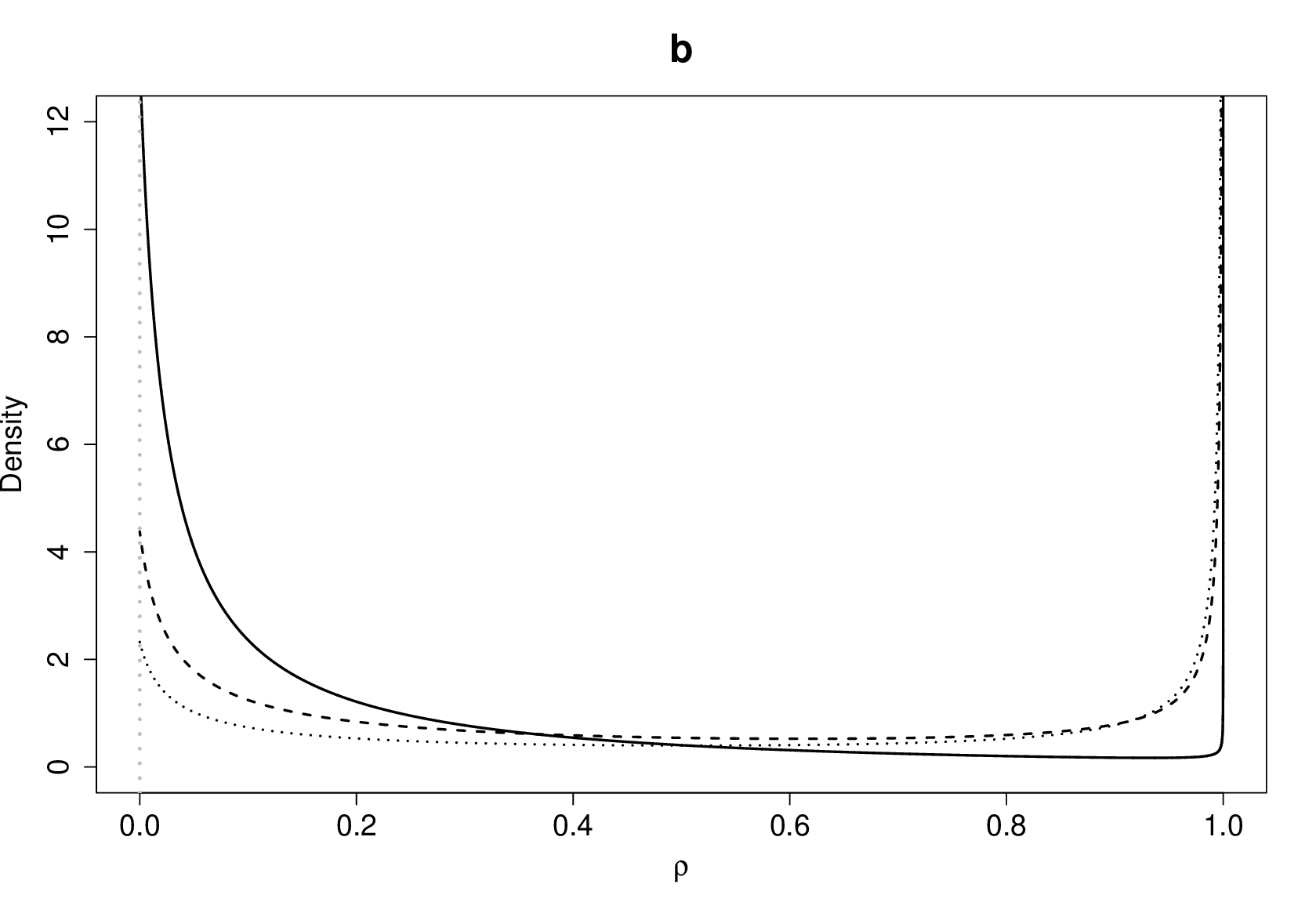}}
\caption{PC prior for $\rho$ in the exchangeable group model - Eq. (\ref{eq:pcprior_rho_exch}) - for different choices of the scaling parameter $\lambda$, according to different priors for the median ICC. The prior is expressed on the scale of the distance from the base model (panel a) and the scale of the original parameter $\rho$ (panel b). The base model is at $\rho=0$ (grey dotted line). The exchangeable model here concerns a design with $m=50$ observations within $ n=6$ groups; the grouping factor is \texttt{campaign}, from the motivating example in Section~ \ref{sec:motivating}. }
\label{fig:pcprior_lambda}
\end{figure}

Scaling the PC prior (i.e., defining $\lambda$) following this rule becomes very intuitive: if $a$ is small, $U$ can be thought of as a lower bound on $\rho$; if $a=0.5$, then $U$ is the prior median for $\rho$. In the exchangeable case, because of the interpretation of $\rho$ as the ICC, the user could simply set $U$ and $a$ according to a prior statement on the proportion of total variance explained by the grouping factor; e.g., for $a=0.5$, choice of $\lambda$ translates into eliciting the ``median proportion of variance explained by the grouping factor''. In presence of weak (or no) information, a sensible strategy is to set the median ICC to $0.5$, so to be exactly half way between the two opposite scenarios: residuals are iid ($\rho=0$) \emph{vs} residuals are completely predicted by the grouping factor ($\rho=1$). Figure \ref{fig:pcprior_lambda} shows the PC prior in Eq. (\ref{eq:pcprior_rho_exch}) for different choices of the median ICC. It can be seen that the smaller the median ICC set at prior, the stronger the penalty for deviating from the iid base model.

\section{Application}
\label{sec:application}

We illustrate an application of the proposed models with the data described in Section \ref{sec:motivating}. 
In order to investigate presence of small scale processes within macroinvertebrate communities we propose comparison of different group models by means of the Bayes factor \citep{kass-raftery-1995}. 
\cite{sorbye-fractional-2018} note that, in case the models under comparison are extensions of the same base model, a convenient approach would be to set, for both models, the same prior on the distance from the base model; in this way, the resulting Bayes factor will only respond to the models under comparison, while being insensitive to the choice of priors for the model hyper-parameters. 
In the remaining of this section we provide an example showing the advantages of PC priors connected to Bayes factors and then discuss the results for our case study.

\subsection{Comparing different group models using the Bayes factor}


Consider the reparametrized mixed model $y_{ij} = \alpha +  \bm x_{ij}^{\textsf{T}} \bm \beta+ \theta_{ij}$ as in Section \ref{sec:group}, where $\bm \theta_j$ is the vector of residuals within group $j$. Let us compare the following models for $\bm \theta_j$: 
\begin{itemize}
\item $\mathcal{M}_1$: $\bm \theta^{(1)}_j$ are exchangeable residuals like in Eq. (\ref{eq:group-reparam}); this model has hyper-parameters $\{\sigma^2,\rho\}$: $0 \leq \rho < 1$ is the intraclass correlation and $\sigma^2$ is the variance; 
\item $\mathcal{M}_2$: $\bm\theta^{(2)}_j$ are structured residuals following an OU process like in Section \ref{sec:group_ou}; this model has hyper-parameters $\{\sigma^2,\phi\}$: $\phi>0$ is the lag-one correlation in log scale and $\sigma^2$ is the variance.
\end{itemize}
The Bayes factor (BF) quantifies the strength of evidence of $\mathcal{M}_1$ compared to $\mathcal{M}_2$ by the ratio of the marginal likelihoods (considering $\alpha$ and $\bm \beta$ as known, without loss of generality):
\begin{eqnarray}
\text{BF}({\mathcal{M}_1} ; {\mathcal{M}_2} ) &=& \frac{\pi(\bm y|\mathcal{M}_1)}{\pi(\bm y|\mathcal{M}_2)} \nonumber \\
&=& \frac{\int_{\mathcal{M}_1} \pi(\bm y|\bm \theta^{(1)},\sigma^2,\rho ) \pi(\bm \theta^{(1)}|\sigma^2,\rho) \pi(\sigma^2) \pi(\rho) d\bm \theta^{(1)} d \sigma^2 d \rho}{\int_{\mathcal{M}_2} \pi(\bm y|\bm \theta^{(2)},\sigma^2,\phi ) \pi(\bm \theta^{(2)}|\sigma^2,\phi) \pi(\sigma^2) \pi(\phi) d\bm \theta^{(2)} d \sigma^2 d \phi}. 
\label{eq:BF}
\end{eqnarray}
From Eq. (\ref{eq:BF}) we can see that BF depends on the data $\bm y$ (and their likelihood) and choices made by the user regarding the models to compare, i.e. $\pi(\bm\theta^{(1)}|\sigma^2,\rho)$ and $\pi(\bm\theta^{(2)}|\sigma^2,\phi)$, and the prior on the hyper-parameters of such models, $\pi(\sigma^2)$, $\pi(\rho)$ and $\pi(\phi)$.

In general, the effect of the prior cannot be separated from the effect of the model when the BF is used.  Quantifying evidence of the alternative models while neutralizing the impact of the priors on hyper-parameters like $\rho$ and $\phi$ is desirable; especially because practitioners are interested in learning about model's goodness of fit and often have no prior knowledge to inform the prior on $\rho$ and $\phi$. To achieve this goal, the same degree of uncertainty would need to be encoded in the priors on  $\{\sigma^2, \rho\}$ (for $\mathcal{M}_1$) and the priors on  $\{\sigma^2,\phi\}$ (for $\mathcal{M}_2$). Regarding $\sigma^2$, this can be done directly by choosing the same density $\pi(\sigma^2)$ for both $\mathcal{M}_1$ and $\mathcal{M}_2$. The hyper-parameter $\rho$ and $\phi$ live in different spaces and have different interpretations, hence to encode the same degree of uncertainty is a much more difficult task. The PC priors proposed in this paper provide a solution to this issue as they are defined on a distance scale $d(\cdot)$, common to both $\mathcal{M}_1$ and $\mathcal{M}_2$.  Following \cite{sorbye-fractional-2018} we assume PC priors on $d(\rho)$ (for $\mathcal{M}_1$) and $d(\phi)$ (for $\mathcal{M}_2$) with equal rate $\lambda$; in this way, $\mathcal{M}_1$ and $\mathcal{M}_2$ have the same distance, at prior, from the iid base model. 
This strategy is reminiscent of the concept of \emph{compatible priors}  \citep{dawid-2001}, i.e. priors that share similarities across the compared models, which were proposed to lessen the influence of priors on Bayes factors in an objective Bayes perspective.

\begin{figure}
\centerline{\includegraphics[angle=270, scale=0.3]{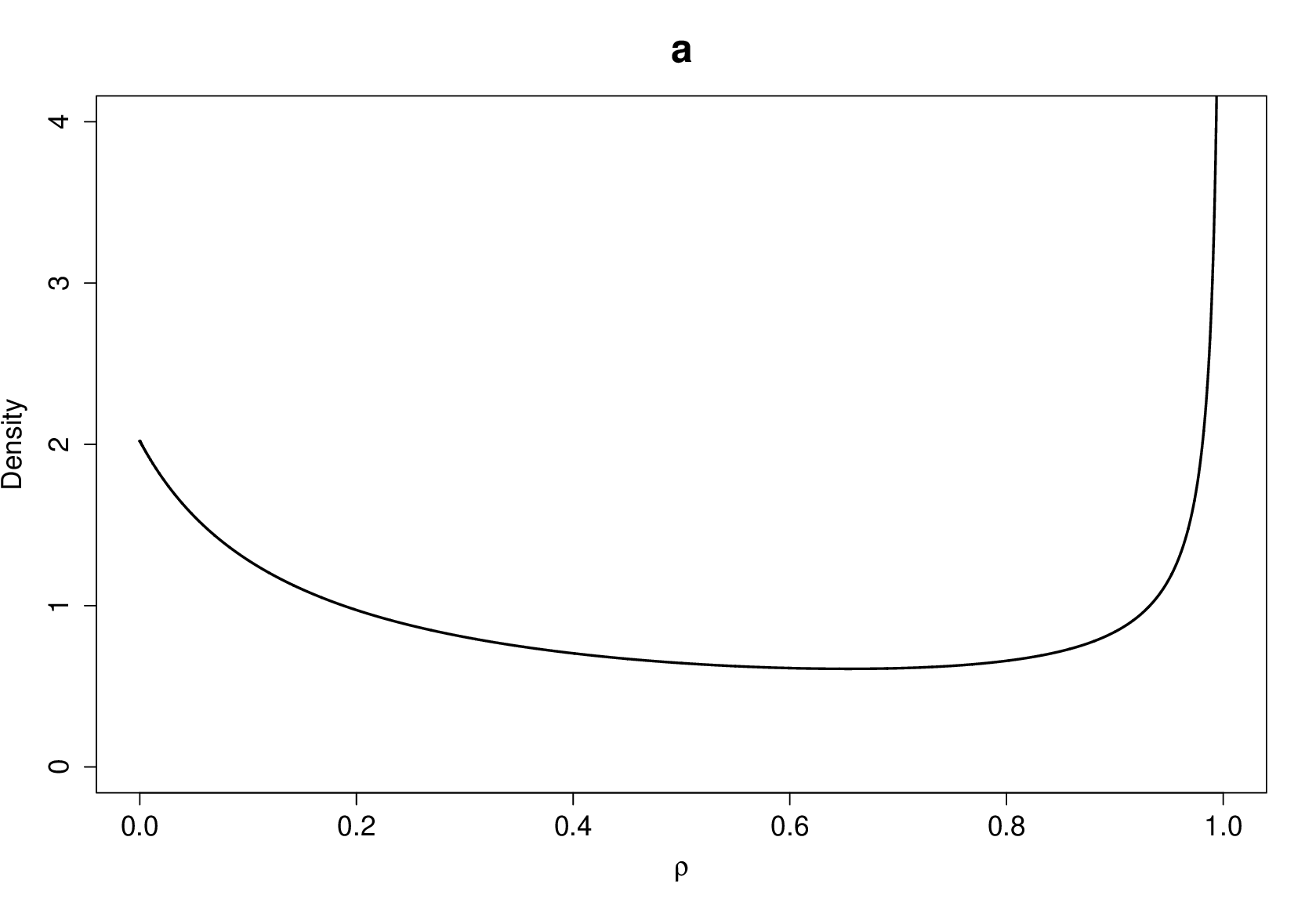}}
\centerline{\includegraphics[angle=270, scale=0.3]{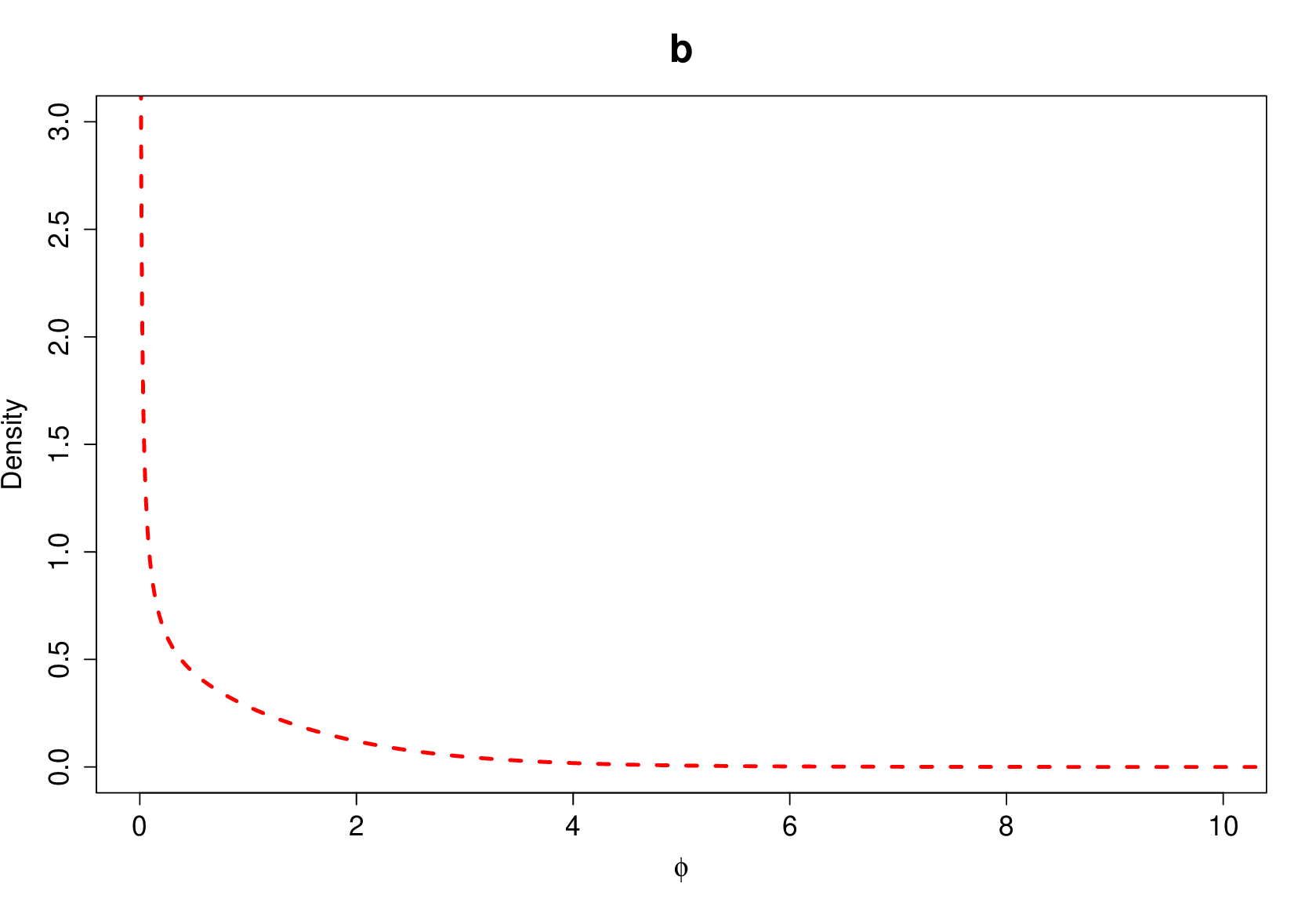}}
\centerline{\includegraphics[angle=270, scale=0.3]{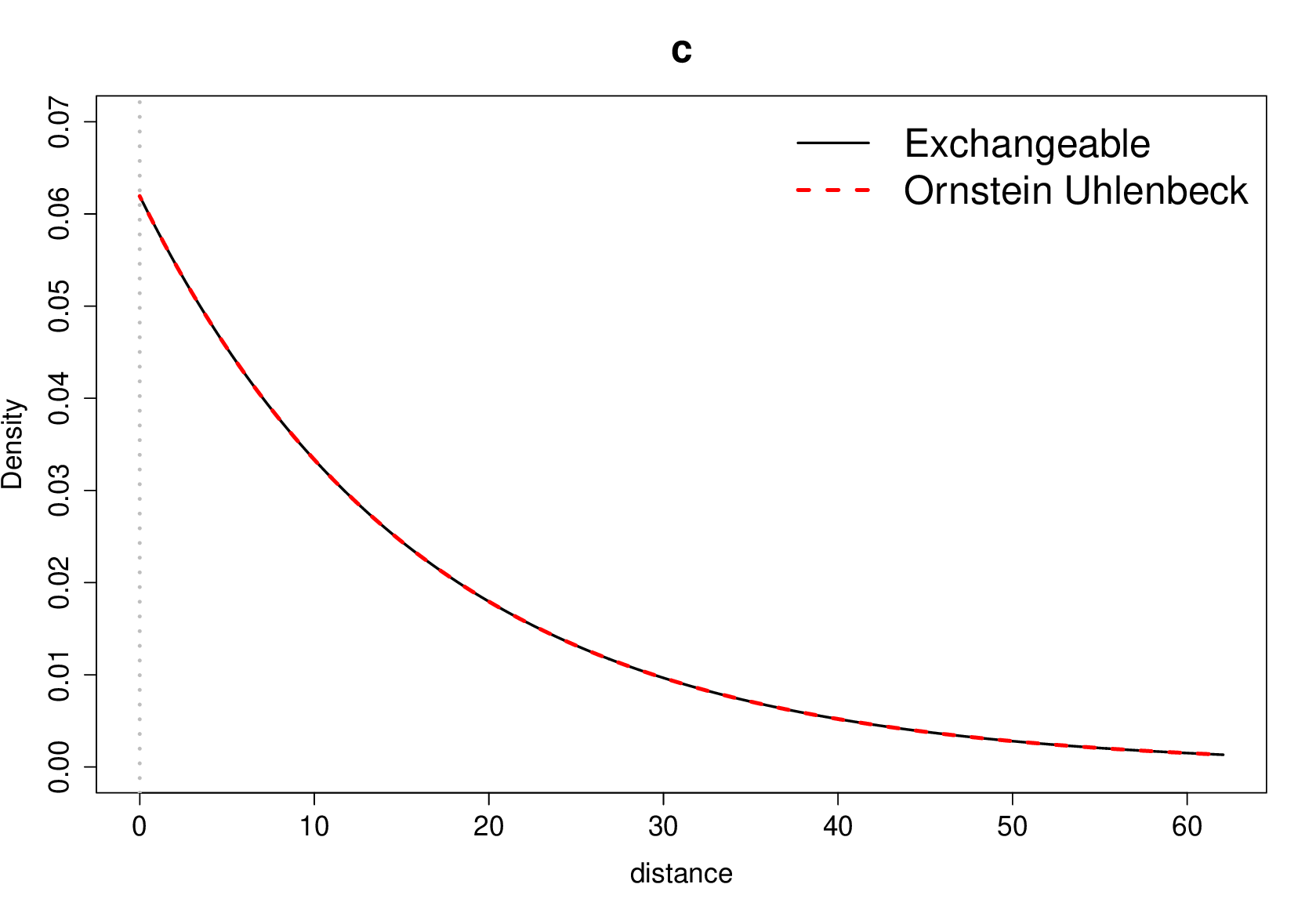}}
\caption{PC priors for two different group models, using the same scaling parameter $\lambda$ (corresponding to a median ICC equal to $0.5$). In panel (a), the PC prior for the correlation $\rho$ of an exchangeable group model. In panel (b), the PC prior for the lag-one correlation $\phi$ (in the log-scale) of a Ornstein Uhlenbeck group model. In panel (c), the two PC priors are displayed in the common distance scale. The grouping factor is  \texttt{transect} for both group models; see Section \ref{sec:motivating}.}
\label{fig:pcprior2_lambda}
\end{figure}

Figure \ref{fig:pcprior2_lambda} helps in clarifying the benefit of using the same $\lambda$ for different group models.  The top panel displays the PC prior on $\rho$ (exchangeable), the central panel shows the PC prior on $\phi$ (OU) and the bottom panel displays both PC priors in the common distance scale. The parameter $\lambda$ is set to have a median ICC equal to $0.5$. While the prior densities materialize differently in the original scales, they are the same in the distance scale (dotted red line and black solid lines are superimposed). 
In our opinion, this shows that specifying group models having the same degree of complexity w.r.t a common base model (or same median ICC) is, in general, a non trivial task that can easily be addressed using PC priors.

\subsection{Results}
We consider the model $y_{ij} = \alpha +  \bm x_{ij}^{\textsf{T}} \bm \beta+ \theta_{ij}$, where $y$ is the log-abundance of macroinvertebrates, covariates $\bm x$ include benthic organic matter (BOM), water depth (P), flow velocity (V), substrate composition (SUB) and season (winter, summer). As a first model we assume residuals $\bm \theta_{j}$ are iid. Visual inspection of residuals for this model indicates substantial structure (results not shown here). In order to understand the nature of such unobserved heterogeneity, we focus on several group models for the residuals, analysing the relevance of grouping factors like \texttt{campaign} and \texttt{transect}. 
We consider three group models for $\bm \theta_{j}$:  
\begin{itemize}
\item $\mathcal{M}_1$: exchangeable residuals within campaign (grouping factor \texttt{campaign}, group model \texttt{exch})
\item $\mathcal{M}_2$:  exchangeable residuals within transect (grouping factor \texttt{transect}, group model \texttt{exch})
\item $\mathcal{M}_3$:  serially correlated residuals within transect (grouping factor \texttt{transect}, group model \texttt{ou})
\end{itemize}
All these models offer a great improvement w.r.t the iid residual model in terms of Bayes factor (figures not shown here) and, perhaps, all three types of structure may be worthy to be included in the final model. However, our aim is not to find the best model for these data, but to illustrate the use of the proposed group models in a real case study. Comparing the above models is beneficial to generate hypotheses on the main sources of unobserved heterogeneity characterizing the ecological community under study. 

The model comparison summaries are reported in Table \ref{tab1}, focusing on three different prior settings regarding the median ICC. From top to bottom, the prior on the median ICC is $0.1$ (more importance assigned to iid residuals), $0.5$ (equal weight to iid and grouping factor) and $0.9$ (more weight assigned to the grouping factor). Within each setting, the three group models have the same prior on the distance scale, thus the marginal likelihoods on the last column represent a fair model comparison tool. Importantly, the Bayes factor here is adopted to fairly compare models that differ as regards to both the grouping factor (e.g. \texttt{campaign} \emph{vs} \texttt{transect}) and the group model (e.g. \texttt{exch} \emph{vs} \texttt{ou}). Bayes factors can be evaluated in the log scale, by taking the difference of the log marginal likelihoods of the compared models. 

All the models were implemented in INLA \citep*{rue-inla}; in Web Appendix 1 an example of the \texttt{R-INLA} code to fit the exchangeable group model is provided.  Regarding the total variance, we adopt the Gumbel$(1/2,\psi)$ type-2 distribution on the precision parameter $1/\sigma^{2}$ (which is the PC prior for the precision of a Gaussian random effect; see \cite{pcprior}, Sec.~3.3, for a proposal on how to choose the scaling parameter $\psi$). 

\begin{table}
\begin{footnotesize}
\caption{Comparison between group models $\mathcal{M}_1$, $\mathcal{M}_2$ and $\mathcal{M}_3$. Results are shown for different prior settings, i.e. prior median ICC equal to $0.1$, $0.5$ and $0.9$. For the OU process, the displayed $\rho$ values refers to the correlation at one meter distance between locations along the transects.}
\label{tab1}
\begin{tabular}{|ll|ccc|c|}
\multicolumn{6}{c}{}\\
\hline
\multicolumn{6}{|l|}{\textbf{prior setting 1}: median ICC $=0.1$ }\\
\hline
 & & \multicolumn{3}{c|}{$\rho$} & \\
grouping factor & group model &    \texttt{0.025q} & \texttt{mean} & \texttt{0.975q} & \texttt{log.mlik}\\  
\hline
\texttt{campaign}&\texttt{exch}&   0.152  &0.3   &   0.492 & -392.522\\
\texttt{transect}&\texttt{exch}&   0.171 &0.282  &    0.418  &-401.053 \\
\texttt{transect}&\texttt{ou}&      0.036   &     0.109   &     0.207   &  -412.727 \\
\hline
\multicolumn{6}{c}{}\\
\hline
\multicolumn{6}{|l|}{\textbf{prior setting 2}: median ICC $ =0.5$}\\
\hline
& & \multicolumn{3}{c|}{$\rho$} & \\
grouping factor & group model &    \texttt{0.025q} & \texttt{mean} & \texttt{0.975q} & \texttt{log.mlik}\\
\hline
\texttt{campaign}&\texttt{exch}&    0.176& 0.331     &  0.52 & -392.848\\
\texttt{transect}&\texttt{exch}&    0.184 & 0.3  &    0.439 & -401.147\\
\texttt{transect}&\texttt{ou}&       0.040   &     0.116    &    0.216  &  -413.779 \\
\hline
\multicolumn{6}{c}{}\\
\hline
\multicolumn{6}{|l|}{\textbf{prior setting 3}: median ICC $=0.9$ }\\
\hline
 &&  \multicolumn{3}{c|}{$\rho$} & \\
grouping factor & group model &     \texttt{0.025q} & \texttt{mean} & \texttt{0.975q} & \texttt{log.mlik}\\
\hline
\texttt{campaign}&\texttt{exch}&   0.191 & 0.349   &    0.538 & -393.208 \\
\texttt{transect}&\texttt{exch}&    0.186 & 0.303 &      0.443   & -401.639 \\
\texttt{transect}&\texttt{ou}&      0.040   &     0.117      &  0.218    &  -414.439  \\
\hline
\end{tabular}
\end{footnotesize}
\end{table}

\subsubsection{Exchangeability within campaign \emph{vs} exchangeability within transect}
We first compare $\mathcal{M}_1$ against $\mathcal{M}_2$, to assess whether residuals are more correlated within campaigns than within transects.
From Table \ref{tab1} we see there is clear evidence that the most relevant grouping factor is campaign; the difference of the log marginal likelihoods is around $9$ units ($401-392$) in favour of exchangeability within campaign, translating into a Bayes factor of around $8100$, i.e. very strong evidence according to the categories proposed by \cite{kass-raftery-1995}. 
Finally, the posterior mean for the correlation within campaign/transect is roughly the same, i.e. around $0.3$. All these results are stable for varying priors on the median ICC.

\subsubsection{Serial correlation  within transect \emph{vs} exchangeability within transect}
We have seen that there is more correlation in the residuals belonging to the same campaign, rather than to the same transect. This does not mean that correlation along transect is not the case. The goal of the study is to assess presence of small scale interactions between organisms. Even if \texttt{transect} has proved to be less important than \texttt{campaign}, it is worth to investigate serial correlation along transects, the latter being a surrogate of small scale interactions between organisms. We then compare $\mathcal{M}_2$ against $\mathcal{M}_3$, where the latter is an OU process on the transect. From Table \ref{tab1} we see there is clearly more evidence in favour of the model implying exchangeability, than the one implying serial correlation along transects; the difference of the log marginal likelihoods is around $11$ units ($413-401$), translating into a Bayes factor of around $60000$, i.e. very strong evidence. The posterior for $\rho$ in the OU case refers to the correlation at one meter distance between locations along the transects; this correlation is around $0.11$. All results are stable for varying priors on the median ICC. 

In conclusion, the presence of small scale interactions remains an open question that should be investigated further. More generally, conclusive evidence for spatially structured unobserved heterogeneity needs to be investigated via models including all relevant random effects simultaneously. We stress that the model comparisons presented above are intended as an exploratory tool, at a further stage an ecologist might consider as a first model the one including exchangeable random effects within campaign, then adding complexity on top of it. 

\medskip 

\begin{figure}
\centerline{
\includegraphics[angle=270,scale=0.45]{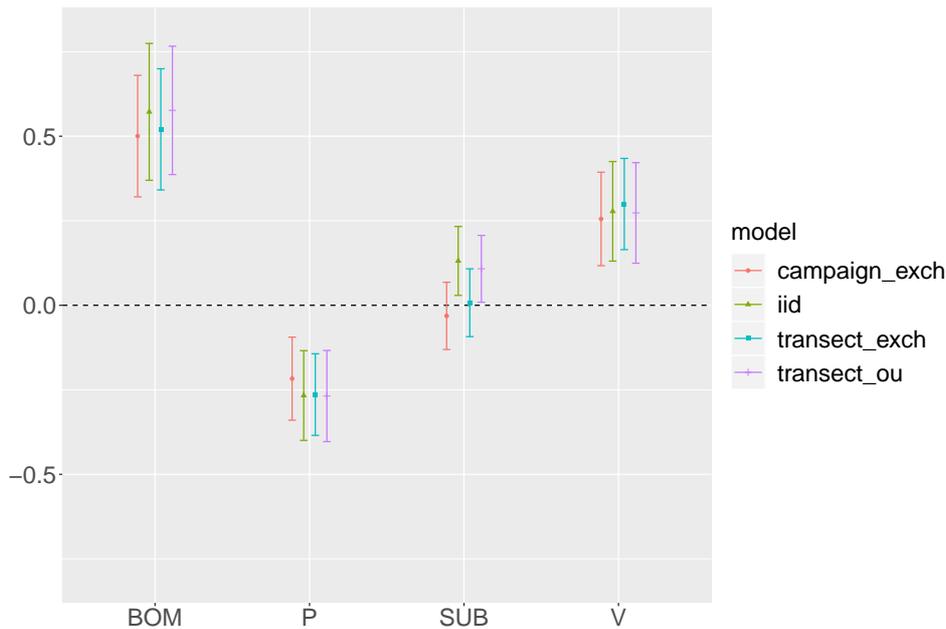}}
\caption{Credible intervals for the covariates (BOM, P, SUB, V) effects, for different group models on the residuals.}
\label{fig:fixed}
\end{figure}

In Figure \ref{fig:fixed} we explore covariates effects for the considered group models and the iid case too. All models track the responses of macroinvertebrates to environmental covariates roughly in the same way, except for the covariate substrate composition (SUB). It can be seen that SUB is found to be significant under the iid model and the OU within transect group model, but it is not under an exchangeable group model accounting for within campaign/transect correlation. This points out the importance of model selection for ecologists, in order to avoid type I errors and misinterpretation of the evidence in the data.

\section{Discussion}
\label{sec:discussion}

Starting from the the one-way anova case, which assumes within group residuals to be exchangeable, we presented group models encoding different assumptions on the residual structure. All these models are an extension of the linear regression base model (iid residuals). Importantly, the generic group model was parametrized so that only one correlation (hyper-)parameter ($\rho$) is responsible for the shrinkage towards the iid case. 
We then derived an intuitive PC prior for the correlation parameter which is built from general principles and can therefore be applied in general to different types of one-factor mixed models.

In community ecology studies, residual correlation can be linked to the effect of unaccounted abiotic factors or unobserved biotic processes, like interactions among organisms (e.g. competition, predation etc).  
Different structures in the residuals match different ecological interpretations about the nature of the unobserved heterogeneity, hence reliable model comparison tools are needed to compare alternative models on the residuals. \cite{saville-2009} proposed approximated Bayes factors for testing random effects in linear mixed models, exploiting closed-form solutions derived using conjugate priors for the variance components; in contrast to their work, our paper focuses on the restricted class of one-factor mixed models but embraces group models with generic covariance, without being restricted to conjugate priors.
 

In our opinion assuming a PC prior for the correlation parameter $\rho$ gives several advantages to the user/ecologist. First, it ensures that the group model shrinks to the linear regression case, which avoids overfitting and is coherent with the way ecologists think about the random effect component, that is as an additional assumption required only when covariates are not enough to explain variability in the data. Second, the PC prior for $\rho$ is easy-to-elicit given a prior statement on the median ICC. The ICC represents a highly intuitive scale to quantify the distance from the base model. 
Third, since PC priors are invariant over reparametrization, such user-defined-scaling in terms of the ICC can be applied in general to any group model, regardless of the interpretation and the scale of $\rho$ in the group model itself.  
Fourth, we show in the application that there is clear advantage in using these PC priors in a model comparison setting: if the same $\lambda$ is used for all the compared group models, the impact of the prior in the compared marginal likelihoods is the same. For this reason, the Bayes factor becomes a convenient tool to compare alternative mixed models, varying according to both the grouping factor and the group model. 

The proposed approach can be extended to the generalized case of a response variable belonging to the exponential family. By including an iid Gaussian term in the linear predictor (reflecting a measurement error), derivation of the PC prior follows straightforwardly. As future work, other group models will be considered like the Mat\'{e}rn covariance, with the spatial range playing the role of $\rho$. The PC prior for the spatial range can be derived numerically, by calculating the matrix determinant and its derivative in (\ref{eq:pcprior_rho}); to gain computational efficiency, it is convenient to work with sparse precision matrices exploiting the SPDE approach by \cite*{finn-spde}. Finally, an important extension of this work would be to consider more than one factor. This requires working out joint PC priors for more than one correlation parameter, in a group model with nested random effects.

\section*{Acknowledgements}
Daniela Cocchi and Massimo Ventrucci are supported by the PRIN 2015 grant project n.20154X8K23 (EPHASTAT) founded by the Italian Ministry for Education, University and Research. Gemma Burgazzi is supported by the project PRIN NOACQUA – responses of communities and ecosystem processes in intermittent rivers a National Relevant Project funded by the Italian Ministry of Education and University (PRIN 2015, Prot. 201572HW8F). The authors thank Maria Franco Villoria and H\.{a}vard Rue for the stimulating comments received about this work.

\bibliographystyle{plain}
\bibliography{biblio_PCprior_groupmodel}

\begin{thebibliography}{}

\bibitem[Dawid and Lauritzen, 2001]{dawid-2001}
Dawid, A. and Lauritzen, S. (2001).
\newblock Compatible prior distributions.
\newblock In {\em Bayesian methods with applications to sciences, policy and
  official statistics. Proceedings of the 6th world meeting}, page 642.
  International Society for Bayesian Analysis, Office for Official Publications
  of the European Communities.

\bibitem[Finley et~al., 2009]{finley-2009}
Finley, A.~O., Banerjee, S., Waldmann, P., and Ericsson, T. (2009).
\newblock Hierarchical spatial modeling of additive and dominance genetic
  variance for large spatial trial datasets.
\newblock {\em Biometrics}, 65(2):441--451.

\bibitem[Fr{\"u}hwirth-Schnatter and Wagner, 2010]{FruhwirthSchnatter-2010}
Fr{\"u}hwirth-Schnatter, S. and Wagner, H. (2010).
\newblock Stochastic model specification search for {G}aussian and partial
  non-{G}aussian state space models.
\newblock {\em Journal of Econometrics}, 154(1):85--100.

\bibitem[Fr{\"u}hwirth-Schnatter and Wagner, 2011]{FruhwirthSchnatter-2011}
Fr{\"u}hwirth-Schnatter, S. and Wagner, H. (2011).
\newblock Bayesian variable selection for random intercept modeling of
  {G}aussian and non-{G}aussian data.
\newblock In {\em J. M. Bernardo, M. J. Bayarri, J. O. Berger, A. P. Dawid, D.
  Heckerman, A. F. M. Smith and M. West (Eds.)}, pages 165--200. Bayesian
  Statistics 9, Oxford.

\bibitem[Fuglstad et~al., 2018]{Fuglstad:2017}
Fuglstad, G.~A., Simpson, D., Lindgren, F., and Rue, H. (2018).
\newblock Constructing priors that penalize the complexity of {G}aussian random
  fields.
\newblock {\em Journal of the American Statistical Association}.

\bibitem[Gelman, 2006]{gelman-2006}
Gelman, A. (2006).
\newblock Prior distributions for variance parameters in hierarchical models
  (comment on article by {B}rowne and {D}raper).
\newblock {\em Bayesian Analysis}, 1(3):515--534.

\bibitem[Heino, 2013]{heino2013environmental}
Heino, J. (2013).
\newblock Environmental heterogeneity, dispersal mode, and co-occurrence in
  stream macroinvertebrates.
\newblock {\em Ecology and Evolution}, 3(2):344--355.

\bibitem[Kass and Raftery, 1995]{kass-raftery-1995}
Kass, R.~E. and Raftery, A.~E. (1995).
\newblock Bayes factors.
\newblock {\em Journal of the American Statistical Association},
  90(430):773--795.

\bibitem[Klein and Kneib, 2016]{klein-2015}
Klein, N. and Kneib, T. (2016).
\newblock Scale-dependent priors for variance parameters in structured additive
  distributional regression.
\newblock {\em Bayesian Analysis}, 11(4):1071--1106.

\bibitem[Kullback and Leibler, 1951]{kld-1951}
Kullback, S. and Leibler, R.~A. (1951).
\newblock On information and sufficiency.
\newblock {\em The Annals of Mathematical Statistics}, 22:79--86.

\bibitem[Lamouroux et~al., 2004]{lamouroux2004biological}
Lamouroux, N., Dol{\'e}dec, S., and Gayraud, S. (2004).
\newblock Biological traits of stream macroinvertebrate communities: effects of
  microhabitat, reach, and basin filters.
\newblock {\em Journal of the North American Benthological Society},
  23(3):449--466.

\bibitem[Lindgren et~al., 2011]{finn-spde}
Lindgren, F., Rue, H., and Lindstr{\"o}m, J. (2011).
\newblock An explicit link between gaussian fields and gaussian markov random
  fields: the stochastic partial differential equation approach.
\newblock {\em Journal of the Royal Statistical Society: Series B (Statistical
  Methodology)}, 73(4):423--498.

\bibitem[Ovaskainen et~al., 2017]{ovaskainen2017make}
Ovaskainen, O., Tikhonov, G., Norberg, A., Guillaume~Blanchet, F., Duan, L.,
  Dunson, D., Roslin, T., and Abrego, N. (2017).
\newblock How to make more out of community data? a conceptual framework and
  its implementation as models and software.
\newblock {\em Ecology Letters}, 20(5):561--576.

\bibitem[Riebler et~al., 2012]{riebler-2012}
Riebler, A., Held, L., and Rue, H. (2012).
\newblock Estimation and extrapolation of time trends in registry
  data—borrowing strength from related populations.
\newblock {\em Ann. Appl. Stat.}, 6(1):304--333.

\bibitem[Rue et~al., 2009]{rue-inla}
Rue, H., Martino, S., and Chopin, N. (2009).
\newblock Approximate {B}ayesian inference for latent {G}aussian models using
  inte- grated nested {L}aplace approximations (with discussion).
\newblock {\em Journal of the Royal Statistical Society, Series B},
  71(2):319--392.

\bibitem[Saville and Herring, 2009]{saville-2009}
Saville, B.~R. and Herring, A.~H. (2009).
\newblock Testing random effects in the linear mixed model using approximate
  bayes factors.
\newblock {\em Biometrics}, 65(2):369--376.

\bibitem[Simpson et~al., 2017]{pcprior}
Simpson, D., Rue, H., Riebler, A., Martins, T.~G., and S{\o}rbye, S.~H. (2017).
\newblock Penalising model component complexity: A principled, practical
  approach to constructing priors.
\newblock {\em Statistical Science}, 32(1):1--28.

\bibitem[{S{\o}rbye} and {Rue}, 2017]{sorbye-2016}
{S{\o}rbye}, S. and {Rue}, H. (2017).
\newblock {Penalised complexity priors for stationary autoregressive
  processes}.
\newblock {\em Journal of Time Series Analysis}, 38:923--935.

\bibitem[{S{\o}rbye} and {Rue}, 2018]{sorbye-fractional-2018}
{S{\o}rbye}, S. and {Rue}, H. (2018).
\newblock Fractional gaussian noise: Prior specification and model comparison.
\newblock {\em Environmetrics}, 29(5-6):e2457.

\bibitem[Ventrucci and Rue, 2016]{ventrucci-rue-2016}
Ventrucci, M. and Rue, H. (2016).
\newblock Penalized complexity priors for degrees of freedom in bayesian
  p-splines.
\newblock {\em Statistical Modelling}, 16(6):429--453.

\bibitem[Verbeke and Molenberghs, 2003]{verb-2003}
Verbeke, G. and Molenberghs, G. (2003).
\newblock The use of score tests for inference on variance components.
\newblock {\em Biometrics}, 59(2):254--262.

\bibitem[Warton et~al., 2015]{warton-2015}
Warton, D.~I., Blanchet, F.~G., O'Hara, R.~B., Ovaskainen, O., Taskinen, S.,
  Walker, S.~C., and Hui, F. K.~C. (2015).
\newblock So many variables: Joint modeling in community ecology.
\newblock {\em Trends in Ecology \& Evolution}, 30(12):766--779.

\bibitem[Wisz et~al., 2013]{wisz2013role}
Wisz, M.~S., Pottier, J., Kissling, W.~D., Pellissier, L., Lenoir, J.,
  Damgaard, C.~F., Dormann, C.~F., Forchhammer, M.~C., Grytnes, J.-A., Guisan,
  A., et~al. (2013).
\newblock The role of biotic interactions in shaping distributions and realised
  assemblages of species: implications for species distribution modelling.
\newblock {\em Biological reviews}, 88(1):15--30.

\bibitem[Zuur et~al., 2009]{zuur-2009}
Zuur, A., Ieno, E.~N., Walker, N., Saveiliev, A.~A., and Smith, G.~M. (2009).
\newblock {\em Mixed Effects Models and Extensions in Ecology with {R}}.
\newblock Springer, New York.
\newblock ISBN 978-0-387-87457-9.

\end{thebibliography}

\newpage
\appendix
\section{Proofs of results in Section \ref{sec:PCpriors-balanced}}\label{sec:app_proofs}
Recall the definition of PC prior as an exponential distribution on the distance $d(\rho)$, with rate parameter $\lambda$,
\[
\pi(d(\rho)) = \lambda \exp(-\lambda d(\rho)) \quad \quad \lambda >0.
\]
If design is balanced then $m_j=m,\forall j=1,\ldots,n$; recall that $n$ is the number of groups while $m$ is the number of within group observations. In this case, the distance function in Eq. (\ref{eq:distance}) simplifies to
\begin{equation*}
d(\rho) = \sqrt{- n \log\left(|\bm{R}(\rho)|\right)} \quad \quad 0 \leq \rho < 1.
\label{eq:distance_balanced}
\end{equation*}
Fixing $\lambda  =\lambda'/ \sqrt{n}$, the PC prior for $\rho$ results (by the change of variable rule)
\begin{eqnarray}
	\pi(\rho) &=&  \lambda \exp\left(-\lambda d(\rho)\right) \left|\frac{\partial d(\rho)}{\partial \rho}\right| \nonumber\\
	&=&\lambda \exp\left(-\lambda d(\rho)\right)  \left| -\frac{n}{2\sqrt{-n \log(|\bm R(\rho)|)}} |\bm R(\rho)|^{-1} \frac{\partial |\bm R(\rho)|}{\partial \rho} \right| \nonumber\\
	& = & \frac{1}{2}|\bm R(\rho)|^{-1}\left|\frac{\partial |\bm R(\rho)|}{\partial \rho}\right| \frac{\lambda'}{ \sqrt{-\log\left(|\bm{R}(\rho)|\right)}} \exp(-\lambda'  \sqrt{-\log\left(|\bm{R}(\rho)|\right)}).  
\label{eq:pcprior_rho_balanced}
\end{eqnarray}
Below, the PC priors in Eq. (\ref{eq:pcprior_rho_exch}), (\ref{eq:pcprior_rho_ar1}) and (\ref{eq:pcprior_rho_ou}) are derived. In each case, the proof is completed by deriving the analytical expression for the term $\frac{1}{2}|\bm R(\rho)|^{-1}\left|\frac{\partial |\bm R(\rho)|}{\partial \rho}\right|$ and plugging it in (\ref{eq:pcprior_rho_balanced}).

\subsection*{Exchangeable}
\begin{proof}[Proof of Eq. (\ref{eq:pcprior_rho_exch})]
Let us consider the compound symmetric matrix $\bm R(\rho)$ as in (\ref{eq:compound-matrix}), where subscript $j$ is removed as we are working under a balanced design.
\cite*{riebler-2012} showed that 
\[
|\bm R(\rho)| = (1+(m-1)\rho) (1-\rho)^{m-1} \quad \quad 0 \leq \rho < 1,
\]
hence the distance function is equal to $d(\rho) = \sqrt{-n \log\left\{(1+(m-1)\rho) (1-\rho)^{m-1}\right\}}$. The derivative term in (\ref{eq:pcprior_rho_balanced}) is
\begin{eqnarray*}
\left|\frac{\partial |\bm R(\rho)|}{\partial \rho}\right| &=& \left| \underbrace{(m-1)(1-\rho)^{m-2}}_{>0} \left\{ \underbrace{(1-\rho)  - (1+(m-1)\rho)}_{<0} \right\} \right| \label{eq:1}\\
& = & (m-1)(1-\rho)^{m-2}\left\{(1+(m-1)\rho) - (1-\rho)\right\}.  \label{eq:2}
\end{eqnarray*}
After some algebraic steps, we obtain
\[
\frac{1}{2}|\bm R(\rho)|^{-1}\left|\frac{\partial |\bm R(\rho)|}{\partial \rho}\right|  = \frac{m-1}{2}\left(\frac{1}{1-\rho} - \frac{1}{1+(m-1)\rho}\right),
\]
which completes the proof.
\end{proof}

\subsection*{Autoregressive of order one}
\begin{proof}[Proof of Eq. (\ref{eq:pcprior_rho_ar1})]
The PC prior for the lag-one correlation of an AR1 is derived by \cite{sorbye-2016}. Here we extend it to group models having within group correlation matrix $\bm R(\rho)$ as in (\ref{eq:ar1-matrix}). It can be shown that 
\begin{equation*}
\bm R(\rho)^{-1} = \frac{1}{1-\rho^2}\bm P \quad \quad ; \quad \quad \boldsymbol{P}= 
\begin{bmatrix}
  1 & -{\rho} & 0 & \cdots &    \cdots    & \cdots & 0 \\
-{\rho} & 1+\rho^2  & -{\rho} &  \ddots&     &  & \vdots\\
0 &  - \rho  & 1+\rho^2   & -\rho &  \ddots &    & \vdots  \\
   \vdots &   \ddots &  \ddots & \ddots & \ddots  & \ddots  & \vdots  \\
 \vdots  &  & \ddots & -\rho & 1+\rho^2 & -\rho &  0\\
\vdots &  &  & \ddots & -\rho & 1+\rho^2 & -\rho\\
0&  \cdots    & \cdots & \cdots & 0& -{\rho} & 1
\end{bmatrix},
\label{eq:ar1-matrix-app}
\end{equation*}
where $|\bm P| = 1-\rho^2$. Thus the determinant of the AR1 correlation matrix is
\begin{equation*}
|\bm R(\rho)| = \frac{1}{|\bm R(\rho)^{-1}|} = (1-\rho^2)^{m-1} \quad \quad 0 \leq \rho < 1,
\end{equation*}
hence the distance function is equal to $d(\rho) = \sqrt{-n (m-1) \log(1-\rho^2)}$. The derivative term in (\ref{eq:pcprior_rho_balanced}) is
\begin{eqnarray*}
\left|\frac{\partial |\bm R(\rho)|}{\partial \rho}\right| &=& 2 \rho (m-1)(1-\rho^2)^{m-2}.
\end{eqnarray*}
After some algebraic steps, we obtain
\[
\frac{1}{2}|\bm R(\rho)|^{-1}\left|\frac{\partial |\bm R(\rho)|}{\partial \rho}\right|  =\frac{\rho(m-1)}{1-\rho^2},
\]
which completes the proof.
\end{proof}

\subsection*{Ornstein Uhlenbeck}
\begin{proof}[Proof of Eq. (\ref{eq:pcprior_rho_ou})]
This proof follows straightforwardly from the AR1 case, by recognizing that $\phi = -\log(\rho)$, hence $\rho = \exp(-\phi)$. In this case, the determinant is
\begin{equation*}
|\bm R(\phi)| =  (1-\exp(-2\phi))^{m-1} \quad \quad \phi>0,
\end{equation*}
and the distance function is equal to $d(\phi) = \sqrt{-n (m-1) \log(1-\exp(-2\phi))}$.
The derivative term in (\ref{eq:pcprior_rho_balanced}) is
\begin{eqnarray*}
\left|\frac{\partial |\bm R(\phi)|}{\partial \phi}\right| &=& 2(m-1)(1-\exp(-2\phi))^{m-2} \exp(-2\phi).
\end{eqnarray*}
After some algebraic steps, we obtain
\[
\frac{1}{2}|\bm R(\rho)|^{-1}\left|\frac{\partial |\bm R(\rho)|}{\partial \rho}\right|  = \frac{(m-1)\exp(-2\phi)}{1-\exp(-2\phi)},
\]
which completes the proof.
\end{proof}


\section{Example code for an exchangeable group model}\label{sec:app_code}
Assume  data vector \texttt{y} and covariate vector \texttt{x} of length $N=mn$. Observations are clustered in $n$ groups (indexed by $j=1,\ldots,n$), the $j^{th}$ group having $m_j$ observations. Assume the group model for the residuals $\bm \theta$
\[
\bm \theta  \sim \mathcal{N}(\bm 0, \bm \Sigma), \quad \quad \bm \Sigma = \tau^{-1} \text{diag} \{\bm R_1(\rho), \ldots, \bm R_n(\rho)\}
\]
where $\tau = 1/\sigma^2$ (\texttt{prec} in the code) and $\bm R_j$ is the exchangeable correlation matrix, with correlation $\rho$ (\texttt{rho} in the code).  The grouping factor is \texttt{group = rep(1:n,each=m)} (although the code below works for a general unbalanced design).  First, we compute the PC prior for $\rho$, scaling it based on a median ICC equal to 0.5.
\begin{lstlisting}
## distance function
d.exch.rho0=function(rho, m.vec) {
  n = length(unique(m.vec))
  if (length(rho)>1) {
    d = matrix(ncol=length(rho), nrow=n)
    for (i in 1:n){
      mi = sum(m.vec==i)
      d[i,] = -log((1+(mi-1)*rho)*((1-rho)^(mi-1)))
    }
    res = sqrt(apply(d, 2, sum))
  } else{
    d = numeric(n)
    for (i in 1:n){
      mi = sum(m.vec==i)
      d[i] = -log((1+(mi-1)*rho)*((1-rho)^(mi-1)))
      res = sqrt(sum(d))
    }
  }
  res
}
## compute PC prior
theta = seq(-15,15,length.out = 100000)
rhofun = splinefun(theta, exp(theta)/(1+exp(theta)))  # internal scale for rho 
 # set lambda
a = 0.5; U = 0.5  # ICC=0.5
lambda = c(-log(1-a)/d.exch.rho0(U, dat$group)) 
 # the PC prior in the 'd' scale
fexp = function(lambda, d) lambda * exp(-lambda * d) 
dfun = splinefun(rhofun(theta), d.exch.rho0(rhofun(theta), dat$group))
 # the PC prior in the 'rho' scale
prior = splinefun(rhofun(theta),  fexp(lambda=lambda,
                                       dfun(rhofun(theta)))*abs(dfun(rhofun(theta), deriv=1)))
\end{lstlisting}
In \texttt{R-INLA} it is possible to implement the group model either using the \texttt{control.group} feature or building it ``manually'' with \texttt{inla.rgeneric.define()}. For illustrative purposes, we show the \texttt{rgeneric} option; the code below can be adapted to other group models. 
The group model is specified inside the function \texttt{myrgeneric.exch}, by coding the precision matrix (\texttt{Q}), the log-prior (\texttt{log.prior}) and the normalizing constant (\texttt{log.norm.const}). The normalizing constant of $\pi(\bm \theta|\tau,\rho)$, for the exchangeable case, is (assuming  a generic unbalanced design)
\[
\sum_{j=1}^n \left\{\frac{m_j}{2} \log{\tau} - \frac{1}{2} \left[ \log(1+(m_j-1)\rho) + (m_j-1) \log(1-\rho)\right]\right\}.
\]
\begin{lstlisting}
library(INLA)
## within j-th group precision matrix (exch. group model)
precision.exch = function(m, rho, tau=1){
  denom <- (rho-1)*((m-1)*rho+1)
  Q = matrix(rho, m, m)
  diag(Q) = -((m-2)*rho+1)
  (tau/denom) * Q
}
## rgeneric
myrgeneric.exch = function (cmd = c("graph", "Q", "mu", "initial",
                                    "log.norm.const", "log.prior", "quit"),
                            theta = NULL)
{
  interpret.theta = function() {
    return(list(prec = exp(theta[1L]),
                rho = exp(theta[2L])/(1+exp(theta[2L]))))
  }
  graph = function() {
    return(Q())
  }
     #  precision matrix, for general unbalanced design
  Q = function() {
    prec = interpret.theta()$prec
    rho = interpret.theta()$rho
    Q.list = list()
    for (i in unique(group_id)){
      m = length(group_id[group_id==i])
      Q.list[[i]] = build.precision.exch(m=m, rho=rho, tau=prec)
    }
    Q = bdiag(Q.list)
    return(Q)
  }
  mu = function() {
    return(numeric(0))
  }
  log.norm.const = function() {
    rho = interpret.theta()$rho
    val = 0
    for (i in unique(group_id)){
      m = length(group_id[group_id==i])
      val = val - (m/2)*log(2*pi) + (m/2)*theta[1L] - 0.5*(log(1+(m-1)*rho)+(m-1)*log(1-rho))       
    }
    return(val)
  }
  log.prior = function() {
    prec = interpret.theta()$prec
    val = inla.pc.dprec(prec,u.prec,alpha.prec,log=T)+theta[1L] +
      log.prior.rho.int(theta[2L])
    return(val)
  }
  initial = function() {
    ntheta = 2
    return(rep(1, ntheta))
  }
  quit = function() {
    return(invisible())
  }
  if (is.null(theta)) theta = initial()
  val = do.call(match.arg(cmd), args = list())
  return(val)
}
\end{lstlisting}
Finally, the INLA call:
\begin{lstlisting}
sdres = 1 # set PC prior for residual total st.dev (approx 1)
mymodel.exch = inla.rgeneric.define(model=myrgeneric.exch, 
                     u.prec=sdres/0.31, alpha.prec=0.01,
                     build.precision.exch = precision.exch,
                     group_id = group,
                     log.prior.rho.int = splinefun(theta, log(prior(rhofun(theta))*abs(rhofun(theta, deriv=1)))))
res = inla(y ~ x + f(id, model=mymodel.exch),
                  data = list(y=y, x=x, id=1:N),
                  family="gaussian",
                  control.family = list(
                    hyper = list(prec = list(
                      initial = 12,
                      fixed = TRUE))),
                  verbose=FALSE)
\end{lstlisting}
\end{document}